# Sul rapporto tra scienza e filosofia
## (alcune riflessioni di un fisico matematico)

C. Lo Surdo

Laboratori Nazionali di Frascati dell'INFN

È opinione diffusa che il drammatico divario oggi esistente tra l'approccio filosofico e quello strettamente scientifico alla conoscenza del mondo fenomenico dipenda soprattutto dal progressivo spostarsi di quest'ultima verso regioni sempre più lontane dalla comune esperienza quotidiana: dalla esperienza cioè non mediata e amplificata da strumenti osservativi abbastanza sofisticati. Benché rifletta una situazione reale, questo giudizio è a mio avviso corretto solo in parte: nel senso che la distanza tra la forma mentis del filosofo e quella dello scienziato è *strutturale*, essendosi resa definitivamente palese da quando la scienza ha cominciato a distinguersi nettamente da quella filosofia in cui era stata fino ad allora immersa – all'incirca a cavallo tra il XVI e il XVII Secolo (e per inciso, quasi esclusivamente in Europa). Il fenomeno ha prodotto una significativa diaspora di potenziali energie filosofiche verso altre attività di ricerca: innanzitutto verso la logica matematica e la filosofia della scienza, e in secondo ordine verso alcune altre particolari scienze positive (ad esempio, verso la psicologia della percezione, oppure la psicologia differenziale, o addirittura le scienze socio-economiche, ecc.). Benché relativamente ridotte, le forze filosofiche continuano poi le loro attività tradizionali lungo percorsi più o meno divergenti da quelli scientifici; e molto lascia prevedere che la divaricazione tra i due orientamenti sia destinata ad ulteriormente allargarsi.

Il nucleo di queste considerazioni mi è stato recentemente riproposto dalla scoperta di una frase nel I° Libro della vichiana "(La) scienza nuova" (Sez. II, C. XXXIX), cui ho rimesso mano dopo la lunga pausa che me ne separava dagli anni del liceo. La frase in questione recita:

«La curiosità, proprietà connaturale all'uomo, che partorisce la scienza, all'aprire che fa della nostra mente la maraviglia, porta questo costume: ch'ove osserva straordinario effetto di natura, come cometa, parelio, o stella di mezzodì, subito domanda che tal cosa voglia dire o significare.»

G. B. Vico (1668-1744) fu una delle menti più aperte e lungimiranti del suo tempo, filosofo, giurista, e soprattutto antropologo ante litteram. Egli visse circa una generazione dopo Newton (1642-1727), e definitivamente dopo Galileo (1564-1642); ma non è chiaro se avesse familiarità con le opere dei due scienziati mentre attendeva a "La scienza nuova" (pubblicata tra il 1720 e il 1729).

Tuttavia avrebbe potuto conoscerle, visto che il 1° Dialogo galileiano (quello che portò il suo autore agli arresti domiciliari) è del 1632, e che i Principia newtoniani apparvero nel 1687; e che sia l'uno che gli altri circolavano largamente tra gli eruditi dell'epoca. Se così fosse, sulla sola base della frase appena riportata si potrebbe concludere che il filosofo napoletano non trasse grande insegnamento da quegli scritti, che sono all'origine della cosiddetta rivoluzione scientifica. Questo confermerebbe la tesi generale dianzi suggerita sul carattere innato delle due diverse attitudini, quella tipica del filosofo e quella tipica dello scienziato. Infatti, qualunque studioso o attento lettore delle opere di Galileo e di Newton non potrebbe non essere colpito in modo negativo da due affermazioni contenute nella sentenza vichiana; e questo, facendo proprie le idee di base sulla scienza che i due grandi uomini ci hanno indicato – pragmaticamente, pagina per pagina dei loro scritti – *e nulla di più recente*. Le affermazioni sono le seguenti:

1): che la curiosità (sottolineo, «che partorisce la scienza») sia specificamente suscitata da eventi «straordinari»;

2): che la domanda naturale che si affaccia alla mente dell'uomo curioso consista nel chiedersi «che tal cosa voglia dire o significare».

Conoscesse o no le opere di Galileo e di Newton, apparentemente Vico non capì che la curiosità di chi guarda il mondo fenomenico con attitudine scientifica, prima di essere attratta da eventi straordinari, lo è a maggior ragione da eventi "ordinari", cioè da eventi che cadono sotto l'esperienza comune. [1] Secondo la leggenda, Newton allargò enormemente l'orizzonte della conoscenza del mondo fisico osservando una mela che cadeva dall'albero (un evento tutt'altro che straordinario), ma riflettendoci sopra in modo nuovo e profondo. In modo simile, la banale goccia d'acqua che si forma alla bocca di un rubinetto mal chiuso sarebbe più che sufficiente a suscitare problematici interrogativi in chi la sapesse osservare con il giusto tipo di curiosità (ovviamente senza possedere le moderne precognizioni di scuola sulla tensione superficiale nei liquidi).

Un'altra prova del fatto che Vico guardava il mondo fenomenico con occhio sostanzialmente estraneo alla scienza positiva (beninteso a quella del suo tempo, ma già tutt'altro che trascurabile), sta poi nel genere di domande che secondo lui l'uomo curioso si pone di fronte ai fenomeni naturali. Se l'italiano di Vico equivale al nostro, la domanda dell'uomo curioso (curioso secondo Vico) è infatti priva di senso, e comunque radicalmente non-scientifica; e tale sarebbe apparsa anche agli occhi di Galileo e di Newton. (Ovviamente quest'ultima non è una verità dimostrata, ma soltanto una presunzione di buon senso.) Da quando esiste "strettamente", la scienza non si pone mai domande di quel tipo.

---

[1] Con pochissime eccezioni, la scienza si occupa di eventi che *non* cadono sotto l'esperienza comune, nel senso approssimativamente definito più sopra, da non più di due secoli scarsi. Ad esempio, Brahe e Keplero stabilirono le loro celebri leggi osservando il cielo ad occhio nudo; e J. Dalton, il chimico inglese che all'inizio del XIX Secolo scoprì la legge dei rapporti ponderali semplici, si servìva essenzialmente di bilance.

Essa si pone invece, sempre e soltanto, il seguente fondamentale problema: quale è la regola, ammesso che di regola si tratti, in base alla quale un certo evento o circostanza, anche molto ordinario/a, si correla ad un altro evento o circostanza? Per esempio: è regola che le bolle di sapone, o le gocce di grasso che galleggiano alla superficie del brodo, tendono a diventare rotonde (in particolare, tanto più rapidamente quanto più sono piccole)? Oppure: è regola che la decomposizione elettrolitica dell'acqua produce idrogeno ed ossigeno nel rapporto ponderale 1:8? Oppure: è regola che i caratteri ereditari si trasmettono come unità singolarmente distribuite ad ogni generazione? Oppure ancora: è regola che un pattinatore in rotazione attorno all'asse corporeo aumenta la sua velocità angolare se accosta le braccia ai fianchi? E così via, lungo una serie di innumerevoli esempi. Insomma, l'"homo scientificus" (per così onorare il maccheronico) si pone innanzitutto il problema di una presunta "regolarità", sotto le convenienti condizioni, degli eventi che cadono sotto la sua attenzione. E una volta possibilmente accertata la validità della regola sottoponendo le sue osservazioni ad un giudizio statistico, elementare o sofisticato che sia, e quindi acquisita una certa capacità predittiva in base al semplice principio di analogia, egli fa, o tende comunque a fare, un passo fatale. Vale a dire, si chiede: come posso ricondurre la regola in questione, ragionevolmente accertata, ad un principio più generale ed unificante, posto che un tale programma sia sensato?

Questo, e soltanto questo, è il tipo di domande che la scienza post-galileiana [2] si pone. Alla maggior parte dei non-addetti ai lavori, filosofi o no, può sembrare poco. Ma non lo è affatto: apre un insieme sconfinato di interrogativi ben posti, attraverso una scala altrettanto sconfinata di difficoltà.

Se in particolare la regola in questione appare abbastanza universale, e il modello che ne prevede le particolari espressioni è identificato con certezza (almeno, al momento corrente), allora siamo di fronte ad una "spiegazione", ad un elemento di piccola o grande importanza del cosiddetto "sapere esplicativo", che rimarrà valida fino a quando sarà possibilmente falsificata. In pratica senza eccezioni, il surriferito modello si rivela allora *di natura matematica.* [3] Perché? Cosa si nasconde dietro questo fatto a priori ingiustificato? Esso pone due grandi interrogativi. Da un lato, quello del perché il mondo dei fenomeni si riveli così regolare nella sua ultima realtà; e dall'altro, del perché la matematica sia così efficace per descriverlo. (È impossibile non menzionare qui il celebre articolo di E. Wigner: "The Unreasonable Effectiveness of Mathematics in the Natural Sciences", (1960)). Il problema è suggestivo e profondo, ed ha suscitato un ampio dibattito di natura al contempo filosofica e scientifica.

---

[2] Nel seguito ci riferiremo sempre ad essa con il termine "scienza".
[3] L'idea di collegare alla matematica la nozione di spiegazione scientifica risale a Platone: nel suo Filebo, egli dà per primo chiara espressione al fatto che una cosiddetta "scienza" è tale nella misura in cui contiene della matematica. Non è quindi illegittima l'opinione dello storico della matematica M. Kline, secondo cui «la realizzazione suprema della cultura greca consistette nel riconoscimento del valore della matematica nella ricerca scientifica.» (in "Mathematical Thought from Ancient to Modern Times", vol. II, (1972), trad. ital. Einaudi (1991)). L'idea di Platone si ritrova poi tale e quale in Kant: «Una disciplina contiene tanta scienza quanta è la matematica che usa.» (in "Metaphysische Anfangsgründe der Naturwissenschaft", (1786)).

Che la matematica sia sempre presente nei modelli predittivi efficaci del mondo dei fenomeni non è materia di opinione, ma un fatto incontestabile, sostenuto da una straordinaria moltitudine di prove. La barriera che ci separa dalla immensa regione della conoscenza non dominabile da questo strumento è infatti quella della complessità – non necessariamente anche formale [4] – del modello, complessità che quasi sempre implica l'inaccessibilità, o l'instabilità, delle risposte che la matematica a prima vista ci propone, o addirittura la loro inesistenza. Ma questo è ben accettabile: non c'è nulla di irragionevole, infatti, nella incapacità, umana o di un automa programmato allo scopo – o anche nella impossibilità di principio –, di "calcolare efficacemente" entro un dato sistema formale. Senza contare l'incapacità (umana) di modellare in senso logico, cioè di "formalizzare", un oggetto abbastanza complesso e irregolare, quasi sempre incompletamente osservabile.

La naturale conclusione alla quale arriviamo sembra dunque essere la seguente: che la selezione naturale, attraverso i suoi lenti e misteriosi meccanismi, ha prodotto una specie, quella di homo sapiens, capace di inventare modelli predittivi di certe particolari "sezioni" del mondo, la natura dei quali modelli si rivela sostanzialmente matematica; con ciò intendendosi che essi risultano isomorfi a certi sistemi formali inventati dalla specie stessa (anche se non sempre questi ultimi forniscono soluzioni concrete, per ragioni di complessità). Quindi il sapiens avrebbe inventato la matematica giusta; e ciò non può avere fatto che ispirandosi alla sua diretta esperienza del mondo reale. Ma una seria difficoltà sorge anche accettando questa tesi: molto infatti sembra suggerire che, più l'attenzione del sapiens si dirige verso regioni fenomeniche difficili da osservare (cioè osservabili soltanto per mezzo di strumenti diagnostici molto raffinati), e delle quali egli certamente non aveva avuto esperienza nel corso della sua storia evolutiva, e più la *sua* matematica, spesso inventata con anticipo di decenni sulle applicazioni, può funzionare! [5]

Insomma, la matematica pre-esisterebbe nella regolarità ultima del mondo; una conclusione che si configura come una sorta di nuovo platonismo, in cui al mondo delle idee si sostituisce un carattere recondito, e molto peculiare, del mondo reale. Ora, se è vero, come è vero, che non esiste linguaggio più efficace di quello matematico per descrivere ciò che è strutturalmente regolare, allora la misteriosa concomitanza denunciata da Wigner si spiega, anzi diventa quasi inevitabile. Il problema si sposta peraltro, in questo modo, sull'altro interrogativo: perché il mondo è regolare, e regolare "così e così"? Al quale una risposta audace, ma non irragionevole, potrebbe essere: perché un mondo non regolare, o non regolare così e così, non potrebbe esistere. Cominciamo infatti a recepire segnali – non ancora abbastanza chiari, ma significativi – che ci suggeriscono la precarietà di certi "equilibri dell'esistere": forse il mondo fisico è come è perché se fosse diverso non potrebbe esistere.

---

[4] Come è ben noto, esistono modelli matematici formalmente assai semplici, che tuttavia non ci permettono di "prevedere" nel senso che comunemente si dà a questa parola. L'esempio più famoso è quello del sistema differenziale ordinario di Lorenz, a soluzioni caotiche.

[5] L'ipotesi della esistenza del positrone da parte di Dirac (1930-31) è uno degli esempi più noti ed eclatanti di questa evenienza.

Ma possiamo ormai chiudere sul "problema di Wigner" citando la antecedente e non dissimile opinione di Einstein (da una lettera a M. Solovine), come sempre ispirata al più profondo buon senso: «Ciò che ci dovremmo attendere a priori è un mondo caotico del tutto inaccessibile al pensiero. Ci si potrebbe (o dovrebbe) aspettare che il mondo sia governato da leggi soltanto nella misura in cui interveniamo ad ordinarlo mediante la nostra capacità ordinatrice: qualcosa di simile all'ordine alfabetico del dizionario. Invece il tipo di ordine creato ad esempio dalla teoria della gravitazione di Newton ha tutt'altra natura. Sebbene gli assiomi della teoria siano imposti dall'uomo, il successo di una tale costruzione presuppone un alto grado di ordine del mondo oggettivo, cioè qualcosa che non siamo per nulla autorizzati ad attenderci a priori. È precisamente questo il "miracolo" che sempre più si manifesta con lo sviluppo delle nostre conoscenze.» È facile, infine, intuire quanto sia breve il passo che da queste idee porta alla particolare "religiosità laica" di Einstein («credo nel Dio di Spinoza che si rivela nell'ordinata armonia di ciò che esiste, non in un Dio che si occupa delle sorti e delle azioni degli esseri umani.») e di molti moderni cultori delle scienze esatte.

La storia è ricca di episodi di dissenso tra scienziati [6] e filosofi [7] tra loro contemporanei. Devo aggiungere – ma forse è una impressione personale – che quando la disapprovazione proviene dal versante filosofico, essa tende spesso ad assumere toni astiosi; mentre nel caso opposto è tipicamente improntata al distacco e/o al sarcasmo, per quanto dura possa essere nella sostanza. Esemplari in tal senso trovo da un lato il supponente, e inaccettabilmente superficiale, giudizio di B. Croce sul positivismo scientifico e relativi (secondo lui) «pseudoconcetti» [8]; e dall'altro, lo sprezzante aforisma di Feynman sulla filosofia della scienza [9]. Proseguendo in questa esplorazione del dissenso, soprattutto avverso tesi filosofiche, ricorderò che Gauss – già famoso come princeps mathematicorum – criticò aspramente, ma soltanto in forma privata, i filosofi contemporanei che pretendevano di pontificare sulle scienze esatte, nonostante la loro insufficiente o nessuna competenza in materia. In una lettera del 1844, egli cita in proposito Schelling, Hegel e von Essenbeck, aggiungendo: «E anche con lo stesso Kant, spesso non va molto meglio: secondo me, la sua distinzione tra giudizi sintetici e analitici è una di quelle cose che *o cadono nella banalità o sono false* (c.vo dr)». A quel tempo Gauss aveva già una conoscenza matura della geometria non-euclidea, la quale è già di per sé una netta confutazione delle idee di Kant (1724-1804) in materia di spazio e relativi "giudizi sintetici a priori". [10] Ancora, non si può passare sotto silenzio la parola di Einstein: «Sono convinto che i filosofi abbiano sempre avuto un effetto nefasto sul progresso del pensiero scientifico, poiché hanno sottratto molti concetti fondamentali al dominio dell'empirismo, nel quale si trovavano sotto il nostro (dei fisici, ndr) controllo, trasferendoli alle intangibili altezze dell'"a priori" (evidente allusione a Kant, ndr). Infatti, anche se dovesse risultare che il mondo delle idee non può essere dedotto dall'esperienza attraverso mezzi logici ma è, in certo senso, una creazione della mente umana, senza la quale non è possibile alcuna scienza, esso risulterebbe altrettanto indipendente dalla natura delle nostre esperienze quanto lo sono i vestiti dalla forma del corpo umano.» (da "The Meaning of Relativity", (1922)).

---

[6] La parola "scienziato" ("scientist") fu coniata da W. Whevell (1794–1866), che nel 1841 divenne direttore ("Master") del Trinity College di Cambridge. Prima non esisteva, perché non veniva attribuita alcuna qualifica specifica a chi si occupava di scienza a pieno tempo.

[7] Qui e nel seguito, in questo termine non sono inclusi gli epistemologi, o filosofi della scienza; un'altra categoria di studiosi che fu generalmente riconosciuta come tale soltanto nel tardo Ottocento. Va ricordato, se ce ne fosse bisogno, che molti dei più grandi uomini di scienza (da Galileo a Newton, a Maxwell, a Einstein, a Poincaré, a Heisenberg, a Schrödinger, .. – per restare in campo esatto) furono anche penetranti ed autorevolissimi epistemologi.

[8] «Il conoscere storico è l'unico dotato di validità teoretica.» Purtroppo, la nostra cultura nazionale deve ancor'oggi molta della sua debolezza a così esimie sciocchezze.

Per quanto possa meravigliare, ancora un secolo fa esistevano circoli di pensiero in cui si coltivava Kant come inventore dei giudizi sintetici a priori, e addirittura come assertore di una sorta di presunta "euclideità soggettiva" dello spazio fisico. Sembra non ci si rendesse conto, in quegli ambienti, che una tesi di fondo della "Critica della Ragione pura" (1781) – che nella conoscenza del mondo fisico «non è la mente che si conforma alle cose, ma piuttosto le cose alla mente.» [11] – scade in concreto in un oscuro e gratuito schema psicocentrico, ogni traccia utilizzabile del quale è praticamente scomparsa dal pensiero positivo posteriore. Per ridurre all'osso la questione: nel linguaggio kantiano, la geometria fisica è "sintetica", mentre la geometria matematica è "a priori". Il punto è per contro che nessuna geometria (e a maggior ragione nessuna fisica) è al contempo sintetica e a priori. Kant non lo capì, e credette di aver dimostrato il contrario. Né ha senso tentare di difendere in questo campo il campione della filosofia tedesca con l'argomento che ai suoi tempi le geometrie non-euclidee non erano state ancora scoperte: ciò che è qui in gioco è un principio generale intrinsecamente fallace, che prescindendo dalle conoscenze empiriche e matematiche

---

[9] «La filosofia della scienza serve agli scienziati più o meno quanto l'ornitologia serve agli uccelli.»
[10] «La nozione di spazio non ha origine empirica, ma è una inevitabile necessità del pensiero.» E ancora: «Le proposizioni della geometria sono conosciute sinteticamente a priori e con apodittica certezza.»
[11] H.J. de Vleeschauwer, in un articolo della Encyclopædia Britannica (1974).

effettivamente disponibili investe in modo scorretto tutto il nostro rapporto cognitivo con il mondo fisico. Questo tipo di errore non è rimediabile; e non a caso Kant era dogmaticamente convinto che le sue idee in materia sarebbero state «la base di ogni futura metafisica che voglia presentarsi in forma di scienza». Non fa dunque meraviglia che con tali presupposti Kant non abbia dato alcun contributo all'unico obiettivo di una geometria fisica (per tacere della fisica in genere) sperimentalmente e matematicamente sensata: quello di indicare/proporre una rappresentazione delle sue figure rispondente ai criteri di formalizzabilità e di correttezza tipici della conoscenza scientifica (esatta) in senso stretto. Per concludere questi rilievi, va anche sottolineato che, sebbene l'idea dei giudizi sintetici a priori fosse già radicalmente contestabile quando fu formulata (1781) – tanto è vero che alcuni spiriti illuminati se ne resero conto appena ne vennero a conoscenza – la sua accezione spazio-temporale risultò «completamente annichilita dalle scoperte del XX-mo secolo. La teoria della relatività ha infatti mutato le nostre concezioni sullo spazio e sul tempo, rivelando loro aspetti del tutto nuovi, dei quali non vi è traccia nelle forme a priori kantiane dell'intuizione pura.» [12]

Le stroncature della filosofia naturale di Kant, e più in generale il dis-prezzo del contributo della filosofia (o almeno della filosofia occidentale) alla scienza da quando la seconda si è resa autonoma dalla prima, non possono certo vedersi, ormai da tempo, come posizioni critiche granché originali; ben inquadrandosi invece nella percezione di una crescente inadeguatezza della filosofia ad interagire positivamente con il corso del progresso scientifico. Questo diffuso sentire non dovrebbe tuttavia indurre, a mio avviso, il convincimento che una riflessione filosofica sulle scienze esatte – o sulla "scienza" tout court, per ciò che questa parola significhi – sia necessariamente superflua o addirittura fuorviante; ma piuttosto sottolineare che per affrontarla in modo proficuo occorrono, oggi come già occorrevano ai tempi di Kant, strumenti specifici acquisibili soltanto attraverso un severo tirocinio diretto, nonché una sana dose di prudenza. Chi non riconosce queste regole del gioco, e ugualmente pretende di far sentire la sua voce, quasi certamente commette errori o produce banalità; come prova una lunga galleria di episodi più o meno celebrati, ma comunque effimeri, penosi o esilaranti a seconda dell'attitudine di chi li giudica con più matura cognizione.

Ormai da molto tempo, gli scienziati rimproverano ai filosofi che scelgono la scienza come oggetto delle loro meditazioni la troppo frequente mancanza di rigore, di concretezza e addirittura di informazione. Ciò ha prodotto una significativa dose di indifferenza, quando non di franca *insofferenza*, da parte del mondo scientifico nei confronti di quello filosofico. Per quanto possa essere motivato e giustificato, non possiamo che deplorare questo stato di cose: scienza e filosofia dovrebbero ancora avere qualche intersezione non vuota. D'altra parte, sembrano abbastanza isolate voci come quella del grande poeta e pensatore indiano Iqbal (morto nel 1938, e da molti accostato al nostro Dante), quando riconosce che il pensiero filosofico *deve tener conto delle nuove scoperte scientifiche, aggiornandosi continuamente ad esse*.

---

[12] W. Heisenberg, "Physics and Philosophy" (1958), trad. ital. in Il Saggiatore (1961), Cap V. Una critica dello stesso segno negativo Heisenberg riserva al principio di causalità – secondo Kant, altro preteso "giudizio sintetico a priori", vedi loc. cit, Cap V.

La pretesa di molti filosofi di accedere a campi della conoscenza che sono al di là delle loro reali competenze è fenomeno duro a morire. Per citare soltanto due esempi estremi, posso ricordare la manifesta incapacità di comprendere la teoria della relatività speciale da parte di Bergson ("Durée e simultanéité: à propos de la théorie d'Einstein", (1922)) («Dio lo perdoni» fu il laconico commento del "chiamato in causa"); e su un altro piano, e molto più recente, la crudele monografia-beffa di Sokal e Bricmont ("Impostures intellectuelles" (1997)), vera e propria "galleria degli orrori" di certa filosofia sedicente scientifica del Novecento, specialmente francese.

Ricapitolando: la scienza ricerca le regolarità del mondo dei fenomeni, e in particolare le scienze esatte lavorano alla possibile organizzazione di queste regolarità in sistemi che diciamo "formali", e che (fiduciosamente) supponiamo coerenti. Questo presupposto di fondo può indurre l'opinione che il pensiero scientifico si evolva attraverso una successione ordinata di passi, e all'interno di schemi fissi e predeterminati: un'idea spesso confusamente percepita in certi ambienti della cultura cosiddetta "umanistica". Come sa bene chiunque possegga un accettabile livello di informazione sul "fenomeno scienza", la realtà è completamente diversa: perché quella evoluzione è di fatto punteggiata di discontinuità, di "catastrofi", di veri e propri stravolgimenti costruttivi del senso comune che ci deriva dall'esperienza quotidiana. Va poi da sé che tali eventi comportano quasi sempre faticosi (e per alcuni che volenterosamente ci provano, vani) sforzi di comprensione, adattamento, assimilazione.

Gli esempi sono molti, e ne propongo qui appresso una scelta.

1): la transizione dalla geometria euclidea alle geometrie (metriche) non-euclidee (Gauss (1828), Riemann (1854));

2): la confluenza delle categorie separate dello spazio (euclideo) e del tempo nello spazio-tempo pseudoeuclideo, e l'associata sostanziale identità tra materia "inerte" ed energia (relatività speciale, Lorentz-Einstein (1904-05)) [13];

---

[13] Il fatto sperimentale che apparentemente costrinse Einstein a riformulare in senso operazionale le nozioni di spazio e di tempo fu la costanza della celerità della luce nel vuoto, nella classe dei sistemi di riferimento cosiddetti "inerziali" (≡ quelli in cui vale la legge d'inerzia). Alcuni storici della scienza mettono tuttavia in dubbio l'autenticità di questa influenza, perché Einstein non fa alcun cenno agli esperimenti di Michelson e Morley nella sua celebre memoria del 1905 (priva peraltro di bibliografia). Una cosa è comunque certa: che Einstein − e *prima* di lui Lorentz − acutamente preferirono avvicinare la meccanica all'elettromagnetismo piuttosto che il viceversa.

3): il carattere indefinito, di segnatura +,+,+,− (od opposta, a seconda delle convenzioni), della metrica spazio-temporale, con le note conseguenze (dilatazione dei tempi, contrazione delle lunghezze) a fronte di trasformazioni di Lorentz (relatività speciale);

4): lo sconcertante legame tra principio di causalità e velocità della luce considerata come limite insuperabile (nel vuoto; ancora, relatività speciale);

5): l'influenza della materia-energia sulle proprietà (geo)metriche dello spazio-tempo, inteso come varietà 4-dimensionale lorentziana, congruentemente differenziabile, con tensore di Riemann non nullo, e viceversa (relatività generale, Einstein-Hilbert (1915-16));

6): lo scardinamento del principio di identità dell'oggetto considerato, che non è più *l'individuo*, ma la folla di *repliche mentali dell'individuo* (meccanica quantistica, *intrinsecamente* probabilistica, Born-Heisenberg-Schrödinger-Dirac, (ca. 1926)), cui fa riscontro

7): l'impossibilità di principio di osservare un oggetto-individuo senza alterarne lo stato in modo imprevedibile;

8): il fatto che le particelle cariche (per esempio elettroni e positroni) interagiscono emettendo o assorbendo fotoni virtuali *in violazione del principio di conservazione di energia e momento* (elettrodinamica quantistica, una teoria anticipata da Dirac (1926) e portata a compimento da Feynman-Schwinger-Tomonaga (ca. 1940));

9): la scoperta che la simmetria fondamentale delle equazioni della cromodinamica quantistica corrisponde alla simmetria del gruppo unitario speciale 3-dimensionale (SU(3)), le cui tre dimensioni sono identificate con i "colori" dei *quarks*;

10): il fatto che i sistemi formali finitamente assiomatizzabili e coerenti sono *essenzialmente* incompleti, e la conseguente impossibilità di garantire la coerenza di un sistema formale non elementare mediante procedure "leali" (Gödel, (1931));

11): la comparsa, in certe applicazioni alla fisica, di logiche diverse da quella naturale (logiche "paracoerenti", logica "quantistica").

Oppure, su altri versanti della scienza positiva:

12): la teoria della evoluzione delle specie attraverso la selezione naturale (Darwin (1859));

13): la scoperta dell'acido desossiribonucleico come codificatore universale dell'informazione genetica nella trasmissione dei caratteri ereditari (1943), nonché quella della doppia elica (Crick e Watson (1953)), con il conseguente avvio delle basi concettuali della genetica;

14): l'infondatezza scientifica − dal punto di vista genetico − dell'esistenza di identità razziali nell'uomo anatomicamente moderno.

L'elenco, che si è voluto interrompere qui, potrebbe continuare a lungo, e parecchie delle scelte che vi figurano scambiate con altre. Esso dà comunque un'idea della crescita del potere

410
predittivo (nel senso dell'implicazione "se …allora") della scienza negli ultimi due secoli circa. Nonostante il lungo cammino percorso, siamo tuttavia ancora ben lontani da una forse utopistica "teoria del tutto": altre e più dissacranti acrobazie mentali ci attendono certamente. Per nominare il problema più importante, sul quale i fisici teorici si accaniscono da quasi mezzo secolo: quali e quanto profonde modifiche delle presenti teorie fisiche saranno necessarie per finalmente celebrare il connubio tra relatività generale e teoria quantistica, quello che custodisce il segreto della cosiddetta "gravità quantistica"?

In conclusione: se c'è un contrassegno della scienza, quello è sì la capacità di stupirci (ricordiamo la "maraviglia" della sentenza vichiana, almeno in questo ovviamente corretta), ma anche la tenace volontà di superare lo stupore perseguendo la crescita del nostro potere predittivo mediante la creazione di teorie fenomenologiche, in particolare di teorie fisico-matematiche. Oltre a quel contrassegno, stabile e universale, la scienza ha accolto ed accoglie un unico dogma: che tutto può alla stretta occorrenza [14] essere rimesso in discussione quando sono in gioco i fondamenti primi. Sebbene offra una divertente replica del paradosso del barbiere (può il dogma stesso essere rimesso in discussione?), è ugualmente chiaro cosa esso significhi. L'essere preparati a radicali mutamenti logici e conoscitivi, e l'accettarli cercando di adattarvisi se necessario, sono presupposti costanti lungo il laborioso percorso del sapere esplicativo. Cediamo un'ultima volta la parola ad Einstein: «Agli occhi dell'epistemologo sistematico, il vero scienziato può apparire come un opportunista privo di scrupoli.» Che è come dire: "Nella scienza non vi è posto per i pre-giudizi (≡ giudizi precedenti l'esperienza, o a priori)". Quanti tra i filosofi che hanno riflettuto e riflettono sul mondo dei fenomeni (e talvolta anche tra gli stessi scienziati) possono vantare una *totale assenza di pregiudizi*?

---

[14] Sottolineo questo "all'occorrenza". Se non è criticamente filtrato, il culto del nuovo come valore in sé può essere pernicioso quanto quello del vecchio. Il troppo stroppia sempre!